# A Fast and Simple Algorithm for Detecting Large Scale Structures

**Giovanni C. Baiesi Pillastrini**[1*]

[1] *Sezione di Ricerca Spettroscopia - U.A.I. C/O I.A.S.F. Via del Fosso del Cavaliere 100,  00133 Roma**, Italy*

**ABSTRACT**

**Aims:** We propose a gravitational potential method (GPM) as a supercluster finder based on the analysis of the local gravitational potential distribution measured by fast and simple algorithm applied to a spatial distribution of mass tracers.
**Methodology:** the GPM performs a two-step exploratory data analysis: first, it measures the comoving local gravitational potential generated by neighboring mass tracers at the position of a test point-like mass tracer. The computation extended to all mass tracers of the sample provides a detailed map of the negative potential fluctuations. The most negative gravitational potential is provided by the highest mass density or, in other words, the deeper is a potential fluctuations in a certain region of space and denser are the mass tracers  in that region. Therefore, from a smoothed potential distribution, the deepest potential well detects unambiguously a high concentration in the mass tracer distribution. Second, applying a density contrast criterion to that mass concentration, a central "bound core" may be identify and quantify in terms of memberships and total mass.
**Results:** using a complete volume-limited sample of galaxy clusters, a huge concentration of galaxy clusters has been detected. In its central region, 35 clusters seem to form a massive and bound core  enclosed in a spherical volume of 51 *Mpc* radius, centered at  Galactic coordinates *l* ~ 63°.7, *b* ~ 63°.7 and at  redshift ~ .36. It has a velocity dispersion of 1,183 *Km/s* with an estimated virial mass of  $2.67 \pm .80 \times 10^{16}$ $M_\odot$.
**Conclusions:** the substantial agreement of our findings compared with those obtained by different  methodologies, confirms the GPM as a straightforward and powerful as well as fast cluster finder useful for analyzing large datasets.

[*] *permanent address: via Pizzardi, 13 -  40138 Bologna -  Italy -  email:* gcbp@it.packardbell.org

# 1. INTRODUCTION

The discovery of massive superclusters from intermediate to high redshift is one of the main tasks of the observational cosmology after N-body simulations, based on different cosmological models and different initial conditions, forecasted the probability to find the most massive objects of the Universe within well-defined range of redshift and mass providing an additional test to discriminate between competing cosmological models [1, 2, 3, 4, 5, 6, 7, 8, 9, 10, 11, 12, 13, 14, 15, 16, 17, 18]. In recent years, the hunt to these unexpected structures has been extended to higher and higher redshift using observational and statistical methods.

From the observational point of view, the recent identification of massive galaxy clusters and superclusters at high redshift are generally obtained by the analysis of X-ray emission and Sunyaev-Zeldovich effect signals. At the same time, after large dataset provided by redshift surveys became available, numerous clustering algorithms based on various theories have been proposed to search for large scale structures [19, 20, 21]. The most common ones are the Friends-of-Friends (FoF) and Density Field methods.

The FoF method is very suitable in searching systems of particles in numerical simulations where all particles have identical masses in volume-limited samples and neighboring galaxies, groups and clusters where the key parameter is a fixed or variable linking length. Because galaxy systems contain galaxies of very different luminosity, the FoF method has the disadvantage that objects of different luminosity or mass are treated identically making difficult a clear distinction between poor and rich galaxy systems if their galaxy number density is similar.

The density field methods overcome this problem because luminosities of galaxies are taken into account during the cluster identification process as well as in the determination of their properties. The basic version of the density field method has been obtained by using cell sizes equal to defined smoothing radii [22]. A variant of the density smoothing uses the Wiener Filtering technique to identify superclusters and voids in the 2dFGRS [23]. In this technique, data are covered by a grid whose cells grow in size with increasing distance from the observer. An important refinement to that variant has been obtained by [24] adopting a constant cell size and constant smoothing radius over the whole sample.

The accomplishment of wide–area surveys of galaxies with spectroscopic follow up, such as Sloan Digital Sky Survey (SDSS) [25], allowed to identify superstructures directly from the large-scale galaxy distribution. Recently, from the SDSS–DR7 main and LRG samples [26], the largest catalog of superclusters (SCLCAT hereafter) has been constructed by [27] using the luminosity density field method. The method selects superclusters as overdensities either above local adaptive or fixed thresholds within the luminosity density fields, a technique however, subjects to a certain degree of arbitrariness in the parameter selection [28].

Here, we want to introduce a different approach to the clustering analysis starting from the basic idea that galaxy systems aggregate by following the laws of gravity no matter how different they are and where the gravitational potential is close connected with the matter density field. It is well-known that the evolution of the large scale density perturbations is tightly related to the potential field and the formation of huge scale structures seen in the galaxy distribution as established by the theory of gravitational instability [29]. It follows that over–dense regions arise due to slow matter flows into the negative potential wells so that, the detection of huge mass concentrations can be carried out observing the regions where very deep gravitational potentials originate.

The most prominent methodology in this line of research is the so-called Power-spectrum clustering analysis [30, 31, 32]. It gains information from the spatial distributions of galaxy systems and reliable estimates of density fluctuations as a function of scale. Instead, from the fundamental assumption that the peculiar velocity field can be represented by a gradient of a scalar potential function, led to the so-called "POTENT" method develops by [33, 34]. The essential idea of POTENT is to reconstruct the peculiar velocity field directly from measurements of galaxy redshift and distance (independently). Such a smoothed field is obtained by averaging the radial peculiar velocities with a tensor window function. It then follows that the potential can be derived by taking the line integral of the radial peculiar velocity components along a radial path. A variant which applies the Wienner filter reconstruction method has been recently suggested by [35]. In this case, the reconstruction of the velocity field in a given region can be decomposed in a component induced by the mass distribution within the region and one induced by the outside mass (tidal effect).

These methodologies have a common characteristic: they use galaxy samples where galaxies are taken as tracers of the velocity field-not of the mass. Therefore, to reconstruct the peculiar velocity field one need independent distance measurement for large samples of galaxies. This can be possible within 300-400 Mpc but at intermediate and high redshift where only sparse distance modulus of Supernova and GRBs are available.

On the contrary, our goal is to define a simpler clustering algorithm capable of analyzing large datasets of objects taken as *tracers of mass density field-not of velocity* designed to detect either single huge superstructures or to display contours of projected surface of potential distributions using little computer time consumption. To achieve this goal the selection procedure to detect overdensities must be the same for all distances from the observer. This condition is permitted by the very slow evolution of the spatial distribution of large scale structures. It allows to investigate the distribution of potential fields to explain properties of the matter distribution at the present time. We follow a recent approach used to study star-forming gas cores in an SPH simulation of giant molecular clouds [36, 37]. They define core-finding methods using the *deepest potential wells* to identify core boundaries. Inspired by this simple physics which underlies clustering processes, we propose a new clustering detector to make use of universality of gravitational clustering behaviors in the context of the exploratory data analysis [38]. This idea can be easily adapted to detect overdense regions in the spatial distribution of mass tracers simply looking for deep potential wells within the map defined by the local potential distribution.

What could be the most convenient mass tracer candidate for this particular clustering analysis? We address our preference toward a complete volume-limited cluster sample, for example, extracted from a reliable cluster catalog where we can use well-defined richness-mass relation to estimate cluster masses, rather than a galaxy sample which requires either the assumption of an arbitrary constant mass-to-light ratio or the application of a more sophisticated method based on the broad band colors of a galaxy sample to estimate the individual galaxy mass. The former gives a quick but rough mass estimation appropriated for exploratory data analyses where precise mass determinations are not mandatory. The latter instead, using an appropriate range of ages, metallicities and dust extinctions, can yield fairly reliable masses for galaxies as demonstrated by [39]. Even if such a method largely increases the software complexity and the data elaboration, it has been developed up to high redshift (z~4) by [40] using the GOODS-MUSIC catalog and, recently, by [41 and references therein] using the COSMOS/ultraVISTA survey.

Our proposed method performs a two-step analysis as follows: first, the deepest potential well is determined either analytically either graphically on the map of the local gravitational potential distribution. Second, assuming that the deepest potential field represents the *center* of a cluster concentration, a density contrast criterion developed by [42] is applied in order to quantify in terms of memberships and total virial mass the *bound core* of that concentration.

The paper is organized as follows: in Sect.2 we introduce the gravitational potential-based method. In Sect.3 we perform an exemplification of how the method can be easily applied to a complete cluster sample studying in detail the region where the deepest gravitational potential is measured and describe the assumed density contrast criterion to identify the bound core of the cluster concentration. The results are compared with other catalogs and various issues addressed by our study are discussed. In Sect.4, the conclusions are drawn.

## 2. THE GRAVITATIONAL POTENTIAL-BASED METHOD (GPM)

As described in the previous section, several methods have been applied to identify superstructures, although they are based on different criteria and algorithms. However, they suffer of a common arbitrary nature in clustering selection which depends on the specific choice and tuning of a number of free parameters. In what follows, we introduce a method based on the determination of the local gravitational potential field distribution generated by a complete volume-limited sample of objects which requires a unique fixed free parameter.

For sake of clarity we perform a numerical simulation to demonstrate the close link of the spatial distribution of matter density with the potential distribution adopting a simple toy model defined by:

i) the mass distribution is represented by point-mass tracers placed in a Bravais lattice with periodic boundary conditions where the unit cell is a cube (all sides of the same length and all face perpendicular to each other) with a point-mass tracer at each corner. The unit cell completely describes the structure of the space which can be regarded as a finite repetition of the unit cell. We restrict the present simulation to a cubic lattice with a side length of $10l$ and a side length of the unit cell equal $l$ (for a total of 1,000 corners/points);

ii) each point lies at positions ($x, y, z$) in the Cartesian three-dimensional space (where $x$, $y$ and $z$ are integer multiples of $l$);

iii) all points have the same mass $m$ so that, within the cubic lattice the mass distribution is perfectly uniform.

Now, being gravity a *superposable force,* the gravitational potential generated by a collection of point masses at a certain location in space is the sum of the potentials generated at that location by each point mass taken in isolation. By measuring the local potential at the position of each object taken one at a time as a test-particle, the map of the local potential distribution generated by the spatial distribution of the whole sample will be displayed.

Hence, given $N_{V_j}$ point-masses located at position vectors $d_i$ (from the observer) within a spherical volume $V_j$ of fixed radius $R_V$ centered on a generic test-particle $j$ at position vector $d_j$ from the observer, then the local gravitational potential generated at position vector $d_j$ by the $N_{V_j}$ point masses $m_i$ ($i = 1,....,N_{V_j}$) is given by the conventional formula

$$\Phi_j = -G \sum_{i=1, i \neq j, i \in V_j}^{N_{V_j}} m_i (d_i - d_j)^{-1} \qquad (1)$$

where *G* is the gravitational constant.

Repeating the calculation for each point mass at *jth* position of the cubic lattice, we provide the whole $\Phi_j$ distribution. Note that $\Phi_j$ is always negative, denoting that the force between particles is attractive. According to Eq. (1), the attractive interaction between objects decreases with distance up to 0 when the distance is greater than $R_V$ assumed large enough so that masses outside $V_j$ should have little influence on the potential determination (we will discuss this assumption later). If in the present simulation we assume for simplicity $R_V = l$, for the geometric properties of our Cubic lattice system, each point mass has $N_{V_j} = 6$ nearest neighbors at a distance $l$. Note that point-masses located on external faces, edges and corners of the cubic lattice have 5, 4 and 3 nearest neighbors respectively; then, the measured $\Phi_j$ at their positions are removed for evident "edge" effect. Under these assumptions, the calculation of $\Phi_j$ will turn out constant at each position $j$ i.e., the $\Phi_j$ distribution is perfectly flat without potential wells as expected from a uniform mass distribution.

Now, supposing to modify the number density of our lattice around the central point P(5*l*, 5*l*, 5*l*) adding symmetrically, for example, 6 new point-masses at a distance of *l*/2 from P and rerun Eq.(1). As expected this artificial clump influences locally the potential distribution as can be seen in the contour plot of the $\Phi_j$ distribution of Fig. 1. The potential well is centered on P embedded in a uniform background.

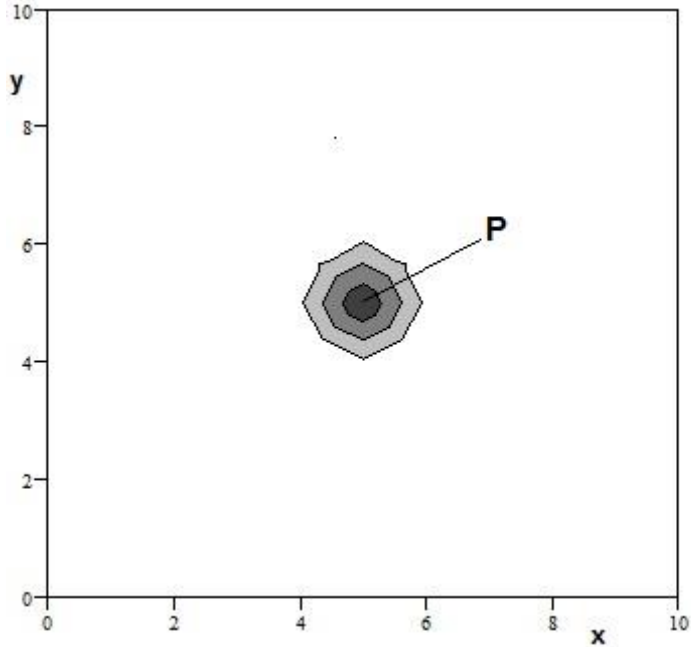

**Fig.1 – Contour plot of the potential distribution in the (x,y)-projection. The potential well centered on P (5*l*,5*l*,5*l*) is shown. The most negative $\Phi_j$ provided by the artificial clump is obtained adding 6 new point mass at *l*/2 around P.**

The astrophysical meaning of the performed simulation can be summarized in a straightforward concept: denser a clump of point-mass tracers embedded in a uniform background is and deeper the generated gravitational potential well will be. *The deepest of the potential wells identifies unambiguously the densest clump of a mass density distribution.* The inverse dependence of $\Phi_j$ on the spatial separation of very close objects as pairs or triplets may provide many local outliers which introduce false signals of very deep wells in the potential distribution. Since we are not interested in detecting small and isolated structures, to overcome this problem we adopt the *mean potential field* $\langle\Phi\rangle_j$ obtained averaging all $\Phi$ measured within each $V_j$. This smoothing process flattens undesirable outliers providing a more realistic evaluation of the potential field and a more reliable signal in detecting huge structures.

The proposed method provides many relevant advantages:

*i*) it enables the identification of clustered structures using an algorithm based on gravity theory where, by knowing positions in space and individual masses of a complete volume-limited sample of mass tracers, local gravitational potentials can be computed from volume density and mapped by contour plots of the potential projected surface;

*ii*) being gravity a long range force, the potential distribution is smoother than the density distribution since the contribution to local potential fields due to small density fluctuations is irrelevant. This is an important property which enables us to constrain overdensities with a clearer physical meaning than, for example, spatial density-based algorithms;

*iii*) it does not require any threshold to be set overcoming the problem of arbitrary density thresholds in the clustering analysis. A property that reduces the complexity of the GPM and enables to build a fast and simple program that can run on a notebook with little computer time consumption and only once.

However, the GPM is data dependent that is, its accuracy in detecting superstructures depends *directly* on the accuracy of the used dataset. In other words, more the dataset defines an accurate representation of mass tracers in *real space*, and more reliable will be the results. It follows that the GPM applied to different datasets obtained by different reconstruction techniques or different selection methods would give different results. This is the main disadvantage of the GPM.

## 3. AN APPLICATION

In what follows we do not perform a rigorous application of the GPM, but only a simplified exemplification to show how it could be applied to a well-defined sample of objects distributed at intermediate redshift.

### 3.1. The data

As demonstrated in the previous section, the GPM can be an accurate clustering algorithm only if the selected sample of objects under study is complete, volume-limited and free of bias effects (selection effect, redshift distortion and so on). For example, if a redshift survey is used to extract a complete volume-limited sample of galaxies, one needs to compare this sample with mock samples derived from cosmological model dependent N-body simulations in order to quantify all bias effects. Such information will be used to implement a reconstruction technique (for example, the POTENT method [36, 37] or the method based on peculiar velocity decomposition [43]) finalized to reconstruct a sample in real space i.e. free of bias effects. Alternatively, one can extract a sample of objects from public catalogs knowing that, normally, they have been tested against mock catalogs to establish the degree of completeness and purity. Since our aim here is to perform an exemplification of how the GPM can be applied, we select an appropriate sample of objects from a public catalog where its accuracy has been tested against N-body simulations.

In gravitational studies, the use of cluster samples overcome some of the problems faced by galaxy samples since clusters are luminous enough for samples to be volume-limited out to large distances and trace the peaks of the density fluctuation field. Therefore, we focalize our analysis on the large scale distribution of galaxy clusters though as point-mass tracers. This assumption is justified by the dynamical state of galaxy clusters which are notoriously relaxed and virialized systems.

The GMBCG optical cluster catalog [44] derived from the SDSS DR7 survey [25] lists each identified cluster by the photometric redshift and coordinates of the brightest cluster galaxy (BCG) as the origin. As mentioned in their work, the Authors have developed an efficient cluster finding algorithm based on the identification of the BCG plus Luminous Red Galaxies (LRG) with a spatial smoothing kernel to measure the clustering strength of galaxies around BCGs. It is well-known that the LRG sample of the SDSS survey provides a sample of intrinsically luminous early-type galaxies selected to impose a passively evolving luminosity cut so as to approach a volume-limited sample. It has been used to trace clusters of galaxies out to $z = .5$ and to provide an enormous volume for the study of large-scale structure. Thus, the SDSS LRG sample maps a comoving volume (expected to reach soon about 1 $h^{-3}$ Gpc$^3$) with a relatively uniform set of luminous, early-type galaxies. They dominate the bright end of the cluster luminosity function and exhibit narrow color scatter colors in the studied redshift range. Therefore, LRG are a very prominent feature of galaxy clusters and thus provide a very powerful means for removing projected field galaxies during cluster detection [45 and references therein]. The success of their method rests on the powerful assumption of existence of a BCG near the bottom of the cluster potential well. This technique provided a catalog of over 55,424 rich galaxy clusters in the redshift range $.1 < z < .55$. The catalog is approximately complete volume-limited up to redshift $z = .4$ and shows high purity and completeness when tested against mock catalogs or compared to analogous cluster catalogs derived from the SDSS survey [46, 47, 48]. The GMBCG has been compared by [48] with their catalog of 55,121 groups and clusters (also derived from the SDSS DR7). The catalog was constructed using the same technique of [44] with a variant: they search for clusters in fields centered on LRG galaxies instead of BCG claiming a substantial agreement and comparable quality with the GMBCG catalog. Then, a comparison has been made with the WHL09 catalog [46] which has been recently updated by [47] (WHL12 hereafter) using the SDSS DR8 survey [49]. Both WHL09 and WHL12 were assembled using a friend-of-friend algorithm applied to luminous galaxies with a linking length of .5 Mpc in the transverse separation and a photometric redshift difference within 3σ along the line-of-sight direction. Similar to the GMBCG, the center is assumed to be the position of the BCG, identified from a global BCG sample, as the brightest galaxy physically linked to the candidate cluster. This new version of the WHL12 was compared with the GMBCG catalog to find the matching rate between them. It has been found that in the redshift range $.05<z<.42$ and richness > 40, about 80% of clusters are matched, but the matching rate decreases abruptly at lower richness. Such a discrepancy, probably due to the application of different clustering algorithms, even if small may lead to different results when the GPM is applied to the same spatial region (see Sect. 3.7.1). Furthermore, relevant discrepancies were found by [50] (see their Fig.5) superimposing a window of the HectoMAP (a map of galaxy clusters selected from *spectroscopic* dataset) on the corresponding part of the GMBCG cluster catalog where many GMBCG clusters do not match the spectroscopic counterpart positions. Evidently, if the GMBCG catalog would suffer of similar bias due to large uncertainties in the photometric data, the results obtained by the application of the GPM may be incorrect.

From the GMBCG cluster catalog, we select a cluster sample belonging to a complete volume-limited spherical shell constrained by the galactic coordinates $0° < l < 360°$ and $60°< b < 82°$ and, a radial thickness of $.1 < z < .4$. Using information on $z$ and the Galactic coordinates $l$ and $b$, for each cluster we determine the comoving radial distance $d$ where the metric is defined by ΛCDM cosmological parameters: $H_0 = 70$ $Km\ s^{-1}Mpc^{-1}$, $\Omega_m = .28$ and $\Omega_\Lambda = .72$.

### 3.2. Simplifying assumptions:

*i)* the GPM measures the local gravitational potential generated by neighboring masses at the position of a point-mass taken as a test-particle on the assumption that the gravitational potential is *time-independent;*

*ii)* The most simplest assumption to connect dark and luminous matter is that galaxy clusters trace the peaks of the underlying matter density field even if the galaxy cluster density is linearly biased with respect to the dark matter density. The exact relationship between the cluster power spectrum and the dark matter power spectrum is well understood theoretically [51], and this relationship or biasing is a function of cluster mass. Then, we may reasonably set the bias factor equal 1 so that fluctuations of the gravitational potential generated by the galaxy cluster distribution also reflect those in the full matter distribution;

*iii)* there is the well-known problem to find a finite solution of $\Phi_j$ for infinite number of gravitating masses. To overcome this problem in our case, we need to assume the form of the spatial distribution of these masses. By considering that at the position of each cluster, the gravitational potential is mainly influenced by its nearest neighbors and much less by other distant masses, we assume a simplified version where a superstructure (clusters of clusters) are approximated by a system of point-like masses (clusters) forming a gravitationally bound system. In this case, we consider such a system as a point-like mass concentrated in its center of mass, which do not interact gravitationally with each other. Further, we assume that this system is surrounded by an empty sphere of fixed radius $R_V$ embedded in a uniform background. Such supposed segregation provides the finiteness of the gravitational potential at any test point inside the sphere but outside where the potential vanishes. Then, we assume $R_V$ = 80 *Mpc* which is capable of a) incorporate the characteristic scale of superclusters i.e. ~ 100-150 $h^{-1}Mpc$ [9]; b) the major share of the gravitational influence exerted on a test cluster by the nearest neighbors (a massive cluster placed beyond $R_V$ has a tiny gravitational influence equivalent to that of a close single galaxy) and c) large enough to avoid the shoot noise error;

*iv*) we consider only the redshift error given in the GMBCG cluster catalog which does not exceed the 10% [39] since the bias affecting high photometric redshift measurements due to relativistic effect has been found negligible on the scales of interest herein [52];

*v*) when $R_V$ overlaps the boundaries of the cluster sample, the measured $\Phi_j$ is automatically ignored in order to minimize false measurements (edge effect).

*vi*) Cluster mass is not directly available from the catalog because only the richness class is provided. Contrary to the cluster mass, the cluster richness may be reliably predicted allowing its use as a proxy for mass (e.g., [53]). However, even if it could be defined precisely for the observational sample under consideration, its use should be taken with care since richness varies depending upon survey characteristics and cluster identification methods. Therefore, we need a richness-mass relation that best fits the GMBCG dataset. From the WHL09 cluster catalog, [54] determined a richness-mass relation for galaxy clusters calibrated using accurate X-ray and weak-lensing mass determinations of a complete sample of clusters defined within .17 < z < .26. The substantial agreement of the richness classification provided by the GMBCG catalog with that of the WHL09 catalog for clusters in common favors the use of that relation to determine the cluster mass of our sample.

## 3.3 Error in the determination of $\Phi_j$

Much of the uncertainty concerning the evaluation of $\Phi_j$ comes from the determination of the cluster mass (which is not observable) as a function of the abundance of the cluster members (richness). The error estimation of $m_i$ (~30%) given by [50 their Eq.4] becomes somewhat arbitrary and poorly determined due to the larger scale ($z$ ~ .4) of our sample. Of course, this assumption is quite questionable because it is well known that the slope of the mass function varies with scales. However, it is worth mentioning that we are dealing with an "exploratory" analysis where a precise mass determination is desirable but not mandatory.

We quantify the error affecting the determination of $\Phi$ using a Monte-Carlo simulation based on the resampling technique [55]. For each data point entering in the calculation of $\Phi$ we assume a Gaussian error distribution with σ as established by [44] for photometric redshifts and [54] for cluster masses. From these distributions we can now randomly sample new data points to estimate the simulated $\Phi$. Repeating this resampling task 10,000 times, we get the distribution of the simulated data from which we can then infer the uncertainty given by the standard deviation. We find that the estimated standard error for $\Phi$ is ~ 32% while the error for $\langle\Phi\rangle_j$ should be reduced by a factor $1/\sqrt{N}$ ( $N$ =number of $\Phi$ measured within $V_j$ ).

## 3.4. Finding cluster concentrations

In Sect.2 we have just outlined that the GPM algorithm shows much less complexity in comparison with conventional clustering algorithms that require parameter tuning and iterative routines. It allows us to implement the GPM algorithm using the Mathcad software to be run on a PC. The program requires only $R_V$ as a free parameter but fixed. The output is obtained running the program just once, either in numerical format as a file where each cluster is identified by its 3D comoving position associated with the calculated $\langle\Phi\rangle_j$ sorted by negative increasing value, either in a graphic format as a Contour plot of the $\langle\Phi\rangle_j$ distribution as a function of Galactic coordinates (see Fig.2).

As stated in Sect.2, when an idealized sample of mass tracers has a perfect uniform distribution, the map of the gravitational potential fields is expected to appear uniform, without wells. Instead, for real sample of mass tracers as galaxy clusters however, contrary to the prediction of the cosmological principle where at large scale deviations from uniformity are expected to be tiny, observations suggest that relevant deviations are present in the form of massive galaxy overdensities and the gravitational influence of such anisotropies should create extended and deep potential wells. It follows that an extreme minimum in the $\langle\Phi\rangle_j$ distribution is the key measurement to detect a huge overdensity in the spatial distribution of our selected cluster sample. This can be clearly seen in the most negative region of the potential distribution of Fig. 2. The isolines of the potential distribution are roughly elliptical for very deep wells, and they become more and more complicated near the zero level as expected for random fields. The point P shows the deepest $\langle \Phi \rangle_j$ = -1.715 x 10$^6$ (Km/s)$^2$ located at Galactic coordinates *l* ~ 62°.7, *b* ~ 63°.1 and *z* ~ .367.

From the visual inspection of the Fig. 2, close to the deepest well at P, appears a secondary deep well indicating another cluster concentration which forms a binary-like system. It lies in the same redshift range of .34 < z < .37 but they centers are separated by more than 180 *Mpc*, prefiguring two distinct cluster concentrations as part of a huge overdensity already detected in the SCLCAT (see Sect.3.7.2). However, in order to follow the aim of this application, we focalize our analysis on the main cluster concentration identified by the deepest well measured at P.

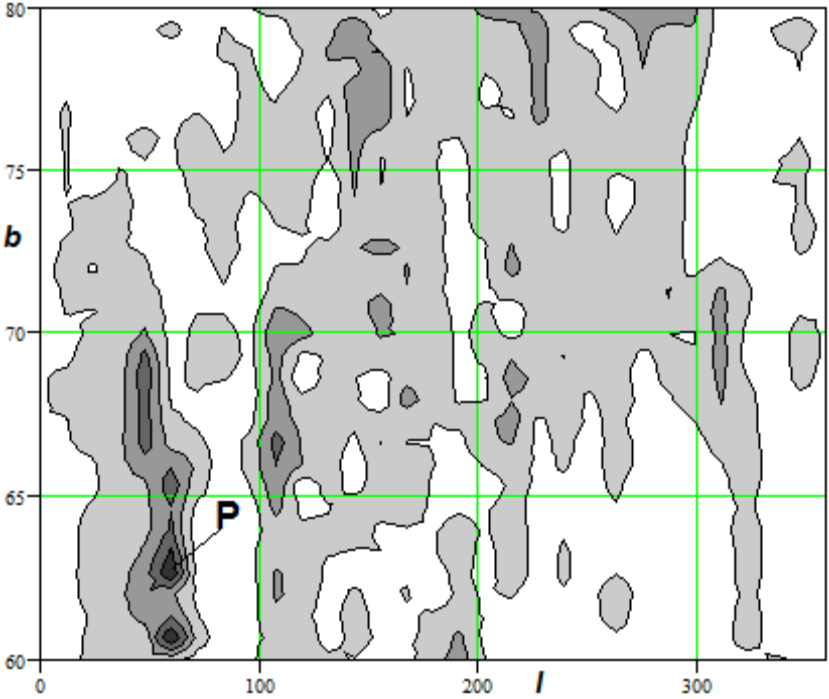

**Fig.2 – Contour plot of the $\langle \Phi \rangle_j$ distribution in the (*l*,*b*)-projection. At the point P (*l=62°.7, b=63°.1, z=.367*), the position of deepest $\langle \Phi \rangle_j$ = -1.715 x 10$^6$ (Km/s)$^2$ is shown. Note also the long chain of cluster overdensities around P (see Sect. 3.7).**

### 3.5. The bound core of the detected cluster concentration

When the GPM is used to search for large "cluster of clusters" (or "concentrations", "superclusters", "associations", "overdensities" of galaxy clusters) as in the present case, one should have in mind that this structures are neither relaxed neither virialized systems. Only after a long dynamical evolution led by gravity, they eventually will virialize. However, by considering the deepest gravitational potential as the maximum density contrast inside a cluster concentration, it is plausible to assume this location as the center of a gravitational *bound* core having in mind that "bound" does not mean fully formed, virialized and clearly separated from each other substructure. To identify unambiguously the memberships of such a hypothetical core is a very difficult task since they have been generally defined by quite arbitrary criteria, mostly on the basis of a statistical algorithms (percolation, FoF, etc.) or caustic technique [56]. Here we adopt the radial density contrast criterion proposed by [42] which assures an accurate process to constrain a massive overdensity with respect to background using simulations in ΛCDM cosmology. This criterion is based on the application of the spherical collapse model to constrain regions enclosed by a spherical shell that eventually evolve into virialized systems. The density contrast that a spherical shell needs to enclose to remain bound to a spherically symmetric overdensity establishes the "lower" density limit for gravitationally bound structures. If $\rho_c$ is the cluster mass density enclosed by the critical shell and $\rho_{bck}$ is the background density (given by $\rho_{crit} \cdot \Omega_m$ where $\rho_{crit}$ is the critical density of the Universe), the mass density enclosed by the last bound shell of a structure must satisfy the density threshold $\delta_c = \rho_c/\rho_{bck}$ = 8.67 (note that in [42], $\delta_c$ = 7.88 due to $\Omega_\Lambda$ = .70 instead of .72 adopted in the present study). All density parameters are determined in unit of M$_\odot$ *Mpc*$^{-3}$. To apply the density criterion, we simply assume that the core of the cluster concentration is defined down to the deepest potential well which it shares with neighboring objects. In this scheme the position of the test cluster where the deepest gravitational potential is measured forms the head of the structure and the center of mass of the densest parts of the cluster concentration. Then, we calculate the density contrast parameter $\delta_{sph,n}$ for *n* concentric spheres with increasing radius until the condition $\delta_{sph,n} < \delta_c$ will be satisfied. Subsequently, we calculate the new center of mass of this sphere and repeat the process iterating until the shift in the center between successive iterations is less than 1% of the radius. With the final center of mass, we identify the angular position, radius and cluster memberships of the bound spherical region corresponding to the core of the cluster concentration. In Fig.3 the radial density contrast profile is apparent. It shows a cusp constrained within 10 *Mpc* radius from the center while in the outer part it gradually fades up to 51 *Mpc* radius where $\delta_{sph,7} = \delta_c$, resulting in an increasingly random cluster distribution. This is the characteristic cusp expected from the collapse of a large structure where the continuous sharpening of the internal mass distribution is reflected in the steepening of its density profile.

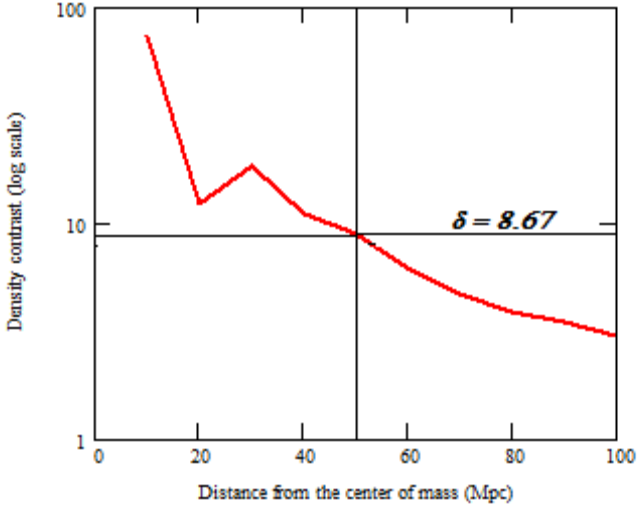

**Fig.3 - Plot of the radial density contrast profile obtained from the application of the clustering algorithm to identify bound core of the cluster concentration. The intersection of the profile with the horizontal line, showing the limit δ = 8.67 of the density contrast criterion, identifies the radius of the critical shell.**

The identified bound core is centered at Galactic coordinate *l*=63°.71 and *b*=63°.72 or J2000 coordinate RA = 14$^h$ 46$^m$ 18$^s$ and Dec = 37° 37' 40" and *z* ~ .36 (hereafter in the text the identified core is labeled as J221°.5+37°.6+036 in decimal degrees + z). J221°.5+37°.6+036 is assembled by 35 clusters enclosed in a sphere of 51 *Mpc* radius. The main properties of its cluster members are summarized in Table 1 as follows: in Col.1, the GMBCG-ID J2000 coordinates in decimal degrees; Col.2 and 3, the photometric redshift and richness class, respectively (these columns are taken from the GMBCG cluster catalog); Col.4, the cluster mass estimation obtained from the richness-mass relation of [54]; Col.5, the measured mean gravitational potential field $\langle \Phi \rangle_j$.

**Table 1. Properties of J221°.5+37°.6+036 cluster members**

| GMBCG ID | z | R | M $10^{14} M_{sun}$ | ⟨Φ⟩ $10^6 (Km\ s^{-1})^2$ |
|---|---|---|---|---|
| GMBCG J219.58196+37.61495 | 0.358 | 13 | 1.43 | -1.443 |
| GMBCG J219.58832+38.09427 | 0.352 | 8 | 0.67 | -1.444 |
| GMBCG J219.64068+36.67152 | 0.350 | 12 | 1.27 | -1.408 |
| GMBCG J219.84475+38.54932 | 0.357 | 13 | 1.43 | -1.428 |
| GMBCG J220.29901+37.53316 | 0.345 | 15 | 1.79 | -1.43 |
| GMBCG J220.34301+36.96806 | 0.362 | 12 | 1.27 | -1.52 |
| GMBCG J220.85463+36.65632 | 0.364 | 22 | 3.24 | -1.602 |
| GMBCG J220.90808+36.73450 | 0.365 | 8 | 0.67 | -1.611 |
| GMBCG J220.91481+35.24239 | 0.353 | 9 | 0.81 | -1.412 |
| GMBCG J221.03525+35.79696 | 0.345 | 17 | 2.17 | -1.391 |
| GMBCG J221.06604+35.95363 | 0.358 | 62 | 16.15 | -1.537 |
| GMBCG J221.11459+35.85098 | 0.358 | 19 | 2.58 | -1.524 |
| GMBCG J221.18593+35.36778 | 0.348 | 8 | 0.67 | -1.426 |
| GMBCG J221.19438+36.16468 | 0.352 | 20 | 2.80 | -1.484 |
| GMBCG J221.31667+36.44627 | 0.343 | 50 | 11.57 | -1.423 |
| GMBCG J221.47047+35.61850 | 0.362 | 18 | 2.37 | -1.554 |
| GMBCG J221.60125+37.98198 | 0.350 | 28 | 4.71 | -1.527 |
| GMBCG J221.62931+38.02956 | 0.356 | 27 | 4.45 | -1.595 |
| GMBCG J221.65575+38.10294 | 0.353 | 49 | 11.22 | -1.596 |
| GMBCG J221.68131+37.99997 | 0.341 | 18 | 2.37 | -1.418 |
| GMBCG J221.73575+37.21649 | 0.352 | 38 | 7.56 | -1.5 |
| GMBCG J221.88634+39.25153 | 0.351 | 14 | 1.61 | -1.459 |
| GMBCG J222.05890+35.19966 | 0.353 | 8 | 6.76 | -1.463 |
| GMBCG J222.13249+35.47027 | 0.355 | 11 | 1.11 | -1.501 |
| GMBCG J222.15362+37.98939 | 0.359 | 56 | 13.79 | -1.633 |
| GMBCG J222.25885+38.00825 | 0.362 | 11 | 1.11 | -1.653 |
| GMBCG J222.44332+37.31708 | 0.367 | 34 | 6.36 | -1.715 deepest |
| GMBCG J222.46480+37.40898 | 0.347 | 13 | 1.43 | -1.555 |
| GMBCG J222.46816+37.58336 | 0.354 | 19 | 2.58 | -1.603 |
| GMBCG J222.49434+38.09762 | 0.356 | 23 | 3.47 | -1.616 |
| GMBCG J222.70127+35.63375 | 0.359 | 9 | 0.81 | -1.566 |
| GMBCG J222.77598+38.55925 | 0.347 | 8 | 0.67 | -1.566 |
| GMBCG J222.92370+38.06561 | 0.356 | 14 | 1.61 | -1.644 |
| GMBCG J223.50442+35.82607 | 0.354 | 26 | 4.20 | -1.61 |
| GMBCG J223.64751+37.60505 | 0.358 | 17 | 2.17 | -1.681 |

In Fig. 4, we show the (*l,z*)-polar projection of our volume-limited cluster sample, displaying the 8,348 galaxy clusters where at their *j* position $\langle \Phi \rangle_j$ has been measured (small black dots). The 35 cluster members of J221°.5+37°.6+036 are also shown (large black dots).

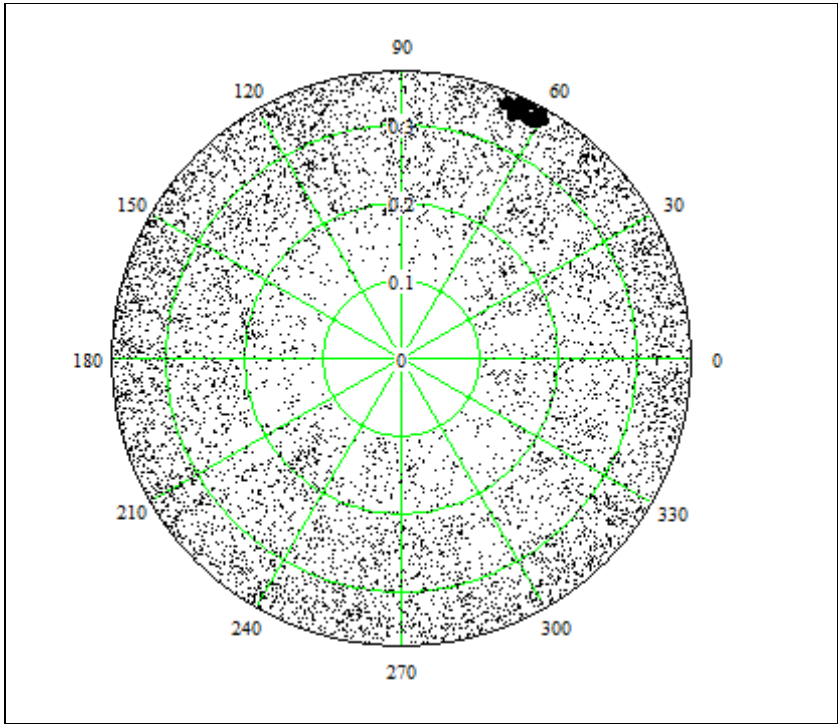

.
**Fig.4 - The (*l,z*)-polar projection of our volume-limited cluster sample. The 35 cluster members of the identified core listed in Table 1 are outlined (large black dots).**

### 3.6. Quantifying the virial mass of J221°.5+37°.6+036

As a first approximation, a mass estimation of J221°.5+37°.6+036 is obtained by summing up the individual cluster masses which yields a value of 1.23 x 10$^{16}$ $M_\odot$. Of course, the total mass is expected to be considerably larger than this, because lower mass as field galaxies are expected to contribute significantly to the total.
The lack of lensing and X-ray data prevents an accurate mass estimation forcing us to use virial mass estimators based on the virial theorem. This estimator requires the assumption of dynamical equilibrium of the system, an assumption quite questionable for large structures as J221°.5+37°.6+036 because many effects like the halo asphericity, secondary infall and lack of the virial equilibrium may heavily bias the result. Nevertheless, needing an approximated mass estimation of J221°.5+37°.6+036, a virial estimator may be used as demonstrated by [57] studying the Corona Borealis supercluster. Assuming that the mass of a supercluster is proportional to a suitable scale velocity and radius, [57] applied different virial mass estimators to the data and tested the results by N-body simulations. They concluded that the error can be restricted to 25-30%. Here, we use the virial estimator $M_{vir} = 3\sigma_v R_{vir} G^{-1}$ (see also [58]) where $R_{vir}$ is estimated as in [59] and the line-of-sight velocity dispersion in the center of mass frame is computed using the prescriptions of [60]. Then, we have found $\sigma_v$ =1,183 *Km/s* and $M_{vir}$ = 2.67±0.80 x 10$^{16}$ M$_\odot$ that is a factor of ~ 2 more massive than the individual mass summation. The estimation of the 1-$\sigma$ error of $M_{vir}$ has been calculated according to the resampling technique described in Sec. 2.3(*vii*). Still according to [57], it is noteworthy that such estimator has the advantage of providing a "conservative" evaluation in case of non-uniform spatial sampling (see also [61], where mass estimations, on average, turn out 20% underestimated and also to [62] which, on the basis of 10,000 Monte Carlo simulations, demonstrated that at least 87% of the virial mass estimations are below the true mass).

An interesting similarity has been found between the properties of J221°.5+37°.6, and those of Shapley concentration (SSC). A detailed study of the SSC was performed by [63] establishing that it is composed of 21 clusters within a sphere of ~ 50 *Mpc* radius and a total mass of 4.4±0.44 x 10$^{16}$ M$_\odot$. In comparison, J221°.5+37°.6+036 shows almost the same extension but is less massive than the SSC in spite of having a more numerous cluster population (actually, looking over the richness class of each cluster listed in Table 1, one can easily recognize that many of them look more like a galaxy *group* rather than a cluster).

### 3.7. Comparison with other catalogs

### 3.7.1 WHL12

In Sect.3.1 we pointed out that two cluster catalogs like GMBCG and WHL09/12, both derived from the SDSS survey, show a certain discrepancy in the matching rate in particular for cluster of low richness class. Such a discrepancy, due to the application of different selection methods, may lead to different results when the GPM is applied. Briefly, to disentangle the issue, we compare the spherical volume covered by J221°.5+37°.6+036 with its counterpart within the WHL12 cluster catalog. We found 28 clusters instead of 35 belonging to J221°.5+37°.6+036 but only 11 of them match exactly a counterpart. Because the discrepancy should be related mainly to difference in richness classification, we compare the richness class of the involved clusters in order to search for systematics. In fact, clusters belonging to WHL12 are generally richer than those of the GMBCG, an expected byproduct due to the application of different selection algorithms to the same dataset. Besides, the WHL12 detects clusters of low richness more or less shifted in space with respect to the GMBCG reducing considerably the matching rate. However, the observed discrepancies are tiny i.e., WHL12 has lesser clusters than those of J221°.5+37°.6+036 but richer, and not so relevant to challenge our finding.

### 3.7.2 SCLCAT

The volume occupied by our cluster sample is part of that studied by [27] to construct the SCLCAT. The first concordance found with the SCLCAT refers to a giant overdensity detected at the lowest density limit of 2.20 and identified as ID=226+034+0359 (RA+Dec+z). It is composed of 6,962 galaxies with a box-diagonal of 2,162 *Mpc h*$^{-1}$ which corresponds to the extended and most negative region appearing around the point P of our Fig. 2 confirming the fair agreement between the two clustering detectors. Moreover, at higher density limits of the SCLCAT, this huge overdensity fragments in several denser structures. We found a tight correspondence of our J221°.5+37°.6+036 with the supercluster ID=222+037+0357 identified at the density limit of 5.40 and composed of 91 galaxies with a box_diagonal of 178 *Mpc h*$^{-1}$. The two structures show a fair agreement between their center positions but J221°.5+37°.6+036 is segregated in smaller and denser volumes. This observed discrepancy unlikely could be attributed to the small difference in the choice of $\Omega_m$ = .27 and $\Omega_\Lambda$ = .73 adopted by the SCLCAT with respect to .28 and .72 adopted in present work, because at z ~ .35, comoving distances differ less than 1%. On the contrary, the different selection method used in the process of identification along with the different hierarchical chain of clustering adopted by [27] i.e., galaxies→superclusters chain in contrast with the clusters→superclusters chain used here, may be effective in smoothing overdense regions. A general problem of the modern hierarchical data clustering algorithms is that clustering quality highly depends on how certain parameters are set. What makes the situation even more complicated is that optimal parameter setting is data dependent. As a result, it may happen that different parts of a given data set require different parameter settings for optimizing clustering quality. In such a context, the application of a global parameter setting to the entire dataset may compromise the final result. For example, [27] report that the weighting factors of their clustering algorithm used to derive the SCLCAT are too high for the highest distances, which cause densities that are too high at the farthest edge of the field confirming that the performance of a clustering selection algorithm may depend on the degree of arbitrariness in the parameter selection. The substantial agreement of our finding compared with the SCLCAT counterpart seem however to confirm the efficiency of the GPM in detecting huge clustered structures in spite of the marked differences characterizing the two selection procedures.

### 3.8 Some remarks

Fig. 2 shows two extended minima in the potential distribution of the cluster sample segregated in tight redshift range between .34 < z <.37 . The sources of these potential wells are two close but well separated massive cluster concentrations. The reason of this mass segregation is presently unclear, however, it may carry important cosmological implications which requires a deeper analysis since it detects an alignment of high density regions in the cluster distribution located in the same redshift range. Such a coherent cluster segregation seems to be in tension with the theoretical expectations of the Cosmological principle which predicts an ever increasing matter homogeneity toward larger scale i.e., in a perfect homogeneous background the gravitational potential fields smooth toward uniformity as well as the local gravitational potentials should tend to a common energy. However, we cannot exclude the hypothesis that the observed mass segregation may be an artifact due to an unknown bias in the data. It is thus necessary to be cautious in interpreting the consequences of our finding in terms of a full 3D cluster distribution since the GMBCG catalog was compiled using photometric redshift and there are not convincing proofs that allow to overcome the suspect that large uncertainties in the measurements may affect our results. As outlined in Sect.2, the demonstration claimed by [50] that high accuracy of spectroscopic redshift with respect to photometric one could be the source of bias in cluster selection, is robust enough to prevents any conclusion about our finding as long as accurate spectroscopic data will confirm it. If so, the discovered J221°.5+37°.6+036 would turn out one of the most massive structures of galaxy clusters detected at intermediate redshift. Besides, it would have a direct cosmological implication since its estimated mass seems to be in tension with the allowable locations predicted in the mass-redshift plane by the ΛCDM model [64].

## 4. CONCLUSIONS

In this work, we have introduced a gravitational potential-based method (GPM) used as a clustering finder. It shows lesser complexity than conventional clustering algorithms that require parameter tuning as FoF, percolation, density thresholds ect., enabling the implementation of a simple program that run only once with very little computer time consumption. It has been applied to a volume-limited cluster sample extracted from the GMBCG cluster catalog in order to investigate the relation between the local potential distribution and volume overdensities in a three-dimensional framework with the aim to identify large cluster concentrations. Following the methodology of the explorative data analysis, the GPM applies a two-step procedure: first, for each sampled cluster, taken one at a time as a test particle, the local gravitational potential generated by neighboring cluster masses is measured at its position. By extending the computation on the whole cluster sample, the GPM provides a detailed map of the negative potential fluctuations where the deepest well detects unambiguously an overdensity within the distribution of the cluster sample. Second, a density contrast criterion or an equivalent one is applied to constrain the bound core, if any, of the identified overdensity.

Mapping the gravitational potential distribution, we have found that the deepest potential well is generated by a huge concentration of galaxy clusters. It has a bound core composed of 35 galaxy clusters enclosed in a sphere of 51 *Mpc* radius located at $l \sim 63°.7$, $b \sim 63°.7$, and redshift $z \sim .36$ with velocity dispersion of 1,183 *Km/s* and an estimated virial mass of $2.67 \pm .80 \times 10^{16}$ $M_\odot$. The substantial agreement of our finding compared with those obtained using different methodologies, confirms that the GPM offers a straightforward, powerful as well as fast way to identify clustered structures from large datasets. Besides, it allows refinements or modifications: for example, if one needs to study the clustering properties of pairs, triplets or small groups, a contour plot of the $\Phi_j$ distribution is more appropriate than that of the mean potential field $\langle\Phi\rangle_j$ distribution used here to detect large structures. On the other hand, by considering the uncertainty affecting the GPM is mainly due to the richness-mass relation adopted here, the major refinement expected to improve its performance is to reduce the scatter between the observable (richness) and the predicted quantity (mass). Such an improvement can be achieved calibrating the richness-mass relation obtained for a much deeper optically selected cluster sample with X-ray or lensing selected counterparts and, measuring how the relation scales towards high redshift.